\documentclass{article}
\usepackage{amssymb}
\usepackage{amsmath}

\input{tcilatex}
\begin{document}

\begin{center}
\bigskip

\bigskip

\bigskip

\textbf{FIXING A PARAMETER OF\ THE GALACTIC HALO: A\ MATHEMATICAL\ MODELLING
BY\ HAMILTONIAN METHOD}

\bigskip

\bigskip

Ruslan Isaev$^{1,a}$, A.A. Potapov$^{2,b}$,\textbf{\ }and K.K. Nandi$%
^{1,2,3,c}$

$\bigskip $

$^{1}$Joint Research Laboratory, Bashkir State Pedagogical University, 3A,
October Revolution Street, Ufa 450000, Russia
\end{center}

$^{2}$Department of Theoretical Physics and Astronomy, Sterlitamak State
Pedagogical Academy, Sterlitamak 453103, Russia

\begin{center}
$^{3}$Department of Mathematics, University of North Bengal, Siliguri 734
013, India

\bigskip

$^{a}$E-mail: subfear@gmail.com

$^{b}$E-mail: potapovaa2008@yandex.ru

$^{c}$E-mail: kamalnandi1952@yahoo.co.in
\end{center}

\bigskip

PACS number(s): 04.50.1h, 04.20.Cv

Key words: Hamiltonian system, metric gravity, galactic halo

\begin{center}
\bigskip
\end{center}

\bigskip

\begin{center}
\textbf{Abstract}
\end{center}

We illustrate how the mathematical modelling of the equations of motion in
terms of autonomous Hamiltonian dynamical system can definitively fix a sign
for an otherwise indefinite sign of a certain astrophysical parameter. \ To
illustrate it, we shall consider the Mannheim-Kazanas-de Sitter solution of
Weyl gravity containing the parameter $\gamma $, which is believed to be
significant in the halo gravity. The strategy we adopt is to calculate the
maximum radius up to which the halo supports stable material circular
orbits. The maximum radius for several observed lenses are calculated for
both signs of $\gamma $, and with the observed value of cosmological
constant $\Lambda $. These lenses (all having approximately the Einstein
radius $R_{\text{E}}\approx 10^{23}$ cm) consistently yield a maximum radius 
$R_{\text{max}}^{\text{stable}}$($\simeq 4.25\times 10^{27}$ cm) \textit{%
inside} the de Sitter radius of the universe only when $\gamma $ is
negative, while a positive $\gamma $ yields $R_{\text{max}}^{\text{stable}}$
always exceeding the de Sitter radius.

\begin{center}
----------------------------------------------------------
\end{center}

\textbf{I. Introduction}

The issue of dark matter (halo matter), arising out of reconciling known
gravitational laws with observed flat rotation curves, is a challenging
problem in modern astrophysics. Observationally, so far, mass of the dark
matter appears to increase with distance in galaxies, but in clusters
exactly the reverse is true, the dark matter distribution actually decreases
with distance. Indeed, for certain dwarfs (such as DD0154) the rotation
curve has been measured to almost 15 optical length scales indicating that
the dark matter surrounding this object is extremely spread out (see, for a
review, Sahni [1]). The total mass of an individual galaxy is still somewhat
of an unknown quantity since a turn around to the $v\varpropto r^{-1/2}$ law
at large radii has not been convincingly observed. On the other hand, dark
matter is attractive and localized on scales less than the cosmological
distances where repulsive dark energy prevails. Therefore, one would expect
that maximum radii of stable circular material orbits in the halo be smaller
than the de Sitter radius of the Universe. Imposing this condition on a
certain model theory with an undetermined parameter $\gamma $, we obtain its
sign unambiguously. This is the idea in this paper.

Although classical Einstein's general relativity theory has been nicely
confirmed within the weak field regime of solar gravity and binary pulsars,
observations of flat rotation curves in the galactic halo still lack a
universally accepted satisfactory explanation. The most widely accepted
explanation hypothesizes that almost every galaxy hosts a large amount of
nonluminous matter, the so called gravitational dark matter, consisting of
unknown particles not included in the particle standard model, forming a
halo around the galaxy. One of the possibility could be that these particles
(WIMPs) encircling the galactic center are localized in a thick shell
providing the needed gravitational field [2]. The exact nature of either the
dark matter or dark energy is yet far too unknown beyond such speculations.
There also exist alternative theories, such as Modified Newtonian Dynamics
(MOND) [3,4], braneworld model [5], scalar field model [6] etc that attempt
to explain dark matter without hypothesizing them. A prominent model theory
is Weyl conformal gravity and a particular solution in the theory is the
Mannheim-Kazanas-de Sitter (MKdS) metric [7] that we shall consider here.

The MKdS solution contains two arbitrary parameters $\gamma $ and $k$ ($%
=\Lambda /3$) that are expected to play prominent roles on the galactic halo
and cosmological scales respectively. While the value of the cosmological
constant $\Lambda =1.29\times 10^{-56}$ cm$^{-2}$ is well accepted, there is
some ambiguity about the sign and magnitude of $\gamma $. From the flat
rotation curve data, Mannheim and Kazanas fix it to be positive and being of
the order of the inverse Hubble length, while Pireaux [8] argues for $%
\left\vert \gamma \right\vert \sim 10^{-33}$ cm$^{-1}$. Edery and Paranjape
[9] fix it to be negative from the gravitational time delay by galactic
clusters while the magnitude is still of the order of inverse Hubble length.
Thus there exist great ambiguity both in magnitude and sign of $\gamma $.

In the present context, we recall the observational fact that massive
neutral hydrogen atoms are executing circular motions in the halo around the
galactic center [10]. The redshifted light from those atoms are measured to
determine their tangential velocities [11]. Therefore it is necessary to
consider massive test particle orbits. However, due to conformal invariance
of the theory, geodesics for massive particles would in general depend on
the conformal factor $\Omega ^{2}(x)$, but here we assume a \textit{fixed}
conformal frame and not considering other conformal variants of the metric.

The purpose of this article is to illustrate how application of a
Hamiltonian system can yield the maximum radius around a galactic center
within which there can exist stable circular orbits of massive test
particles and beyond which the orbits become unstable. The criterion of
stability has been originally suggested by Edery and Paranjape [9] because
it provides a way to the determination of a natural length scale or region
of influence of localized sources in the cosmological setting. The strategy
we adopt here is to frame the geodesic equation in the MKdS solution as a
Hamiltonian system, and based on it, analyze the stability of circular
motion, which then favors a negative $\gamma $.

\textbf{II. Geodesic equations}

The MKdS metric is given by [7,9] (in units $G=1$, vacuum speed of light $%
c_{0}=1$):%
\begin{equation}
d\tau ^{2}=B(r)dt^{2}-\frac{1}{B(r)}dr^{2}-r^{2}(d\theta ^{2}+\sin
^{2}\theta d\varphi ^{2}),\text{ \ }B(r)=1-\frac{2M}{r}+\gamma r-kr^{2},%
\text{\ }
\end{equation}%
where $k$ and $\gamma $ are constants.

Using $u=1/r$, we get the following path equation for a test particle of
mass $m_{0}$ on the equatorial plane $\theta =\pi /2$ (see Appendix A):%
\begin{equation}
\frac{d^{2}u}{d\varphi ^{2}}=-u+3Mu^{2}-\frac{\gamma }{2}+\frac{M}{h^{2}}+%
\frac{1}{2h^{2}u^{2}}\left( \gamma -\frac{2k}{u}\right) ,
\end{equation}%
where $h=\frac{U_{3}}{m_{0}}$, the angular momentum per unit test mass. For
photon, $m_{0}=0\Rightarrow h\rightarrow \infty $ and one ends up with the
conformally invariant equation but without $k$ making its appearance:%
\begin{equation}
\frac{d^{2}u}{d\varphi ^{2}}=-u+3Mu^{2}-\frac{\gamma }{2}.
\end{equation}%
In the Schwarzschild-de Sitter (SdS) metric, such a disappearance has been
noted for long [12] but here we find that it occurs despite the presence of $%
\gamma $ in the metric. The impact of $k$ and $\gamma $ on light bending has
been investigated elsewhere, in Refs.[13-16].

Analysis of dynamical system involves converting the second order equation
into two first order equations [17,18] (see Appendix B). For this purpose,
we introduce the notation%
\begin{equation}
u=x\text{, }y=\overset{.}{x}=\frac{dx}{d\varphi }
\end{equation}%
to reduce Eq.(2) into a pair of first order autonomous system in the ($x,y$)
phase plane%
\begin{align}
\overset{.}{x}& =X(x,y)=y \\
\overset{.}{y}& =Y(x,y)=a+bx+cx^{2}+dx^{-2}+ex^{-3}
\end{align}%
where%
\begin{equation}
a=\frac{M}{h^{2}}-\frac{\gamma }{2}\text{, }b=-1\text{, }c=3M\text{, }d=%
\frac{\gamma }{2h^{2}}\text{, }e=-\frac{k}{h^{2}}.
\end{equation}

First we discuss stability of circular orbits of light, though their
stability is not essentially needed. Recall that even in the Schwarzschild
spacetime, circular light orbits at $R=3M$ are unstable. Nonetheless, it is
instructive to have a look at this aspect in the MKdS solution.

\textbf{III. Dynamical autonomous system: Massless particle motion}

Light motion occurs in circular orbits defined by $R(2+\gamma R)-6M=0$ [see
Eq.(11) below] because $h^{2}\rightarrow \infty $, which implies that $d=e=0$
but $\gamma \neq 0$. The equilibrium points are given by $\overset{.}{x}=0$, 
$\overset{.}{y}=0$, which yield $\left( \frac{-b+\sqrt{b^{2}-4ac}}{2c}%
,0\right) $ and $\left( \frac{-b-\sqrt{b^{2}-4ac}}{2c},0\right) .$ To locate
these points on the real phase plane ($x,y$), we must have $\alpha
^{2}\equiv b^{2}-4ac=1+6\gamma M\geq 0.$ Now $\alpha ^{2}=0\Rightarrow
\gamma =-\frac{1}{6M}$, so the equilibrium points reduce to one single point
given by $P:\left( \frac{1}{6M},0\right) $. For\ $\alpha ^{2}>0$, or $\gamma
>-\frac{1}{6M}$, there are two distinct equilibrium points points $Q_{\pm
}:\left( \frac{1+\alpha }{6M},0\right) $ where $\ \alpha =$ $\pm \sqrt{%
1+6\gamma M}$. Thus $Q_{\pm }$ correspond to two $\gamma -$dependent light
radii $R_{\pm }=\frac{6M}{1\pm \sqrt{1+6\gamma M}}$, which expand as follows%
\begin{equation}
R_{+}=\frac{-1+\sqrt{1+6M\gamma }}{\gamma }\approx 3M+O(\gamma ),
\end{equation}%
\begin{equation}
R_{-}=\frac{-1-\sqrt{1+6M\gamma }}{\gamma }\approx -3M-\frac{2}{\gamma }%
+O(\gamma ).
\end{equation}%
We have from Eq.(23) below%
\begin{equation}
q_{0\pm }=1-\frac{6M}{R_{\pm }}
\end{equation}%
which yields $q_{0+}=-\sqrt{1+6M\gamma }<0$ leading to unstable radius at $%
R=R_{+}$, while $q_{0-}=\sqrt{1+6M\gamma }>0$ showing that $R=R_{-}$ is a
stable radius. The basic constraint (reality condition) for both is that $%
\gamma >-\frac{1}{6M}$. With, say, $\gamma =-7\times 10^{-28}$ cm$^{-1}$
(value inspired by Ref.[9]), the constraint is always satisfied for known
lenses (say, for $M=2.9\times 10^{18}$cm, Abell 2744). From Eq.(9), then we
get the value $R_{-}=2.86\times 10^{27}$ cm, at which there is stability.
However, stability of massive particle circular orbits is more relevant,
which we examine next.

\textbf{IV. Hamiltonian system: Massive particle motion }

The equilibrium points are given by $\overset{.}{x}=0$ and $\overset{.}{y}=0$%
. The equation $\overset{.}{x}=0$ gives $r=R=$ constant, while $\overset{.}{y%
}=0$ gives%
\begin{equation}
h^{2}=-\frac{2MR^{2}+R^{4}(\gamma -2kR)}{R(2+\gamma R)-6M}.
\end{equation}%
The autonomous system (5), (6) can be phrased as a Hamiltonian system as
follows 
\begin{align}
\frac{\partial H}{\partial x}& =-Y\left( x,y\right)
=-(a+bx+cx^{2}+dx^{-2}+ex^{-3}) \\
\frac{\partial H}{\partial y}& =X\left( x,y\right) =y.
\end{align}%
The necessary and sufficient condition for the system (12),(13) to be a
Hamiltonian system, namely, $\frac{\partial X}{\partial x}+\frac{\partial Y}{%
\partial y}=0,$ is fulfilled for all $x$ and $y$. Moreover, $\frac{dH}{%
d\varphi }=0$ and therefore $H\left( x,y\right) =$ constant (independent of $%
\varphi $). Integrating Eqs.(12),(13), we get%
\begin{equation}
H(x,y)=-(ax+\frac{b}{2}x^{2}+\frac{c}{3}x^{3}-dx^{-1}-\frac{e}{2}x^{-2})+u(y)
\end{equation}%
\begin{equation}
H(x,y)=\frac{1}{2}y^{2}+v(x)
\end{equation}%
where $u(y)$ and $v(x)$ are arbitrary functions subject to the consistency
of Eqs.(12) and (13). These two equations will match only if 
\begin{equation}
u(y)=\frac{1}{2}y^{2}+C
\end{equation}%
\begin{equation}
v(x)=-(ax+\frac{b}{2}x^{2}+\frac{c}{3}x^{3}-dx^{-1}-\frac{e}{2}x^{-2})+E
\end{equation}%
where $C$, $E$ are arbitrary constants. The family of Hamiltonian paths on
the phase plane are given by%
\begin{equation}
H(x,y)=\frac{1}{2}y^{2}-(ax+\frac{b}{2}x^{2}+\frac{c}{3}x^{3}-dx^{-1}-\frac{e%
}{2}x^{-2})+G
\end{equation}%
where $G$ is a parameter. It follows that 
\begin{equation}
\frac{\partial ^{2}H}{\partial x^{2}}=-(b+2cx-2dx^{-3}-3ex^{-4})
\end{equation}%
\begin{equation}
\frac{\partial ^{2}H}{\partial y^{2}}=1
\end{equation}%
\begin{equation}
\frac{\partial ^{2}H}{\partial x\partial y}=0.
\end{equation}%
As before, the equilibrium points occur when $X=0$ and $Y=0$, which give the
values of the orbit radius $r=R=$ constant and $h^{2}$ as in Eq.(11). The
quantity determining stability is [17] 
\begin{equation}
q\equiv \frac{\partial ^{2}H}{\partial x^{2}}\frac{\partial ^{2}H}{\partial
y^{2}}-\left( \frac{\partial ^{2}H}{\partial x\partial y}\right) ^{2}.
\end{equation}%
Putting the value of $h^{2}$, we get at the equilibrium points, suffixed by
zero, the following expression%
\begin{equation}
q_{0}=1-\frac{6M}{R}+\frac{R(3kR-\gamma )[R(2+\gamma R)-6M]}{%
R^{2}(2kR-\gamma )-2M}.
\end{equation}%
This expression is more general than that in the Schwarzschild spacetime,
where $k=\gamma =0$. When $q_{0}>0$ at any point $P:(R,0)$, we say that $P$
is a stable center, but it is an unstable saddle point if $q_{0}<0$. When $%
q_{0}=0$, it is an inflection point where the system begins to become
unstable. Thus $q_{0}=0$ gives $R=R_{\text{max}}^{\text{stable}}$, beyond
which the orbits begin to become unstable. There is also a singular radius $%
R=R_{\text{sing}}$ where $q_{0}$ blows up. In all cases we have studied, $R_{%
\text{sing}}<R_{\text{E}}$ for $\gamma $ negative and $R_{\text{sing}}$ $>R_{%
\text{dS}}$ for $\gamma $ positive. In either case, the occurrence of such
singularity is a bit queer but fortunately it occurs at radii not accessible
to gravitational lensing observations.

We shall calculate the values of $q_{0}$ for observed lenses. Due to
advances in technology, mass $M$ and Einstein radius $R_{\text{E}}$ for
several lenses are available. The observed lens masses $M$ are used to
compute the values of $q_{0}$ marching from the Einstein radius $R_{\text{E}%
} $ out to $R_{\text{dS}}$. A meaningful stable radius should lie between $%
R_{\text{E}}$ and $R_{\text{dS}}=\sqrt{\Lambda /3}=1.52\times 10^{28}$cm.
Fig.1 illustrates for one of the observed lenses (Abell 2744, $M=2.90\times
10^{18} $cm, $R_{\text{E}}=2.97\times 10^{23}$cm), which shows that stable
material radii exist for \textit{all} radii $R\geq R_{\text{E}}$ extending
even beyond the dS radius when $\gamma =+7\times 10^{-28}$ cm$^{-1}$, which
seems physically unlikely due to the fact that in the regions beyond dark
matter, repulsive dark energy takes over. In the cosmological region, where
the spacetime becomes time dependent and accelerating, the existence of
circular orbit itself becomes impossible. Also, the singular radius occurs
at $R_{\text{sing}}=8.14\times 10^{28}$cm $>R_{\text{dS}}$. (Fig. 2). On the
other hand, when $\gamma =-7\times 10^{-28}$ cm$^{-1}$, stability is
achieved up to a radius $R=4.25\times 10^{27}$ cm, which is well within $R_{%
\text{dS}}$ and beyond that maximum radius instability begins to appear. In
this case, the singular radius occurs at $R_{\text{sing}}=9.11\times 10^{22}$%
cm $<R_{\text{E}}$ (Fig. 3). Thus the maximum radius up to which stable
material circular orbits are admissible is $R_{\text{max}}^{\text{stable}%
}=4.25\times 10^{27}$ cm for a negative $\gamma $. This limit is not much
sensitive either to the exact value of $\Lambda $ or to other negative
values of $\gamma $ (such as argued by Pireaux [8]). It can be verified
that, for other lens data as well (as listed in Ishak et al. [14], but not
tabulated here), the \textit{maximum }stable radius remains very nearly the
same. We must caution that this $R_{\text{max}}^{\text{stable}}$ is not the
size of the halo because it would be too big; it is just the maximum allowed
radius of stable orbits.

\textbf{V. Conclusion}

Dynamical systems have been in use in gravitation theories, particularly in
the regime of cosmology. The application of the dynamical system approach to
higher order gravity has been shown to be a powerful tool, revealing global
features of curvature quintessence models (see, for instance, [21]), where
the cosmological equations \textit{per se} are posed as a dynamical system
in higher dimensional phase space. Posing the geodesic equations as
dynamical system for studying their stability within the galactic halo
regime also seems interesting in its own right.

From the foregoing analysis, we conclude the following: A finite maximum
stable radius $R_{\text{max}}^{\text{stable}}$ exists within $R_{\text{dS}}$
only when $\gamma $ is negative. Much less is yet conclusively known about
the halo content or size, which has engendered many theoretical models. The
occurrence of singularity in $R$ is actually a defect of the model under
consideration, not of the Hamiltonian method. The method is quite general
and any other metric solution, if available for the halo, can be used
equally. The key physical input we used was that the stable material
(neutral HII) finite circular orbits exist only inside the halo (giving rise
to flat rotation curves [10]) and not outside it. The Hamiltonian analysis
supports such orbits only for a negative $\gamma $. This is the main result
of the paper, which is probably is of some interest in view of the fact that
it potentially decides the long standing ambiguity in the sign of $\gamma $
mentioned earlier.

\textbf{Acknowledgments}

The authors thank Guzel N. Kutdusova for administrative assistance.

\begin{center}
\textbf{Appendix A}
\end{center}

The standard form of geodesic equation is given by 
\begin{eqnarray*}
\frac{d^{2}x^{\mu }}{d\tau ^{2}}+\Gamma _{\nu \lambda }^{\mu }\frac{dx^{\nu }%
}{d\tau }\frac{dx^{\lambda }}{d\tau } &=&0, \\
g_{\nu \lambda }dx^{\nu }dx^{\lambda } &=&d\tau ^{2}.
\end{eqnarray*}%
Using the expression for Christoffel symbols $\Gamma _{\nu \lambda }^{\mu }$%
, it can be easily rewritten in a convenient form as (see, for instance,
[20]):%
\begin{equation}
\frac{dU_{\mu }}{dp}-\frac{1}{2}\frac{\partial g_{\nu \lambda }}{\partial
x^{\mu }}U^{\nu }U^{\lambda }=0,  \tag{A1}
\end{equation}%
\begin{equation}
g_{\nu \lambda }U^{\nu }U^{\lambda }=m_{0}^{2},\text{ \ }  \tag{A2}
\end{equation}%
where $d\tau =m_{0}dp$, $m_{0}$ is the rest mass of the test particle, $%
U^{\lambda }$ ($=\frac{dx^{\lambda }}{dp}$) is the 4-velocity, here $x^{\mu
}=(t,r,\theta ,\varphi )$ and $g_{\nu \lambda }$ is the metric tensor given
by (1). The advantage in redefining the affine parameter $\tau \rightarrow p$
is that the equations (A1), (A2) are now applicable to both massive and
massless particle motion since $dp\neq 0$.

When we construct the ($\mu =2$) $\theta -$ component from (A1), we find
that $\theta $ must be a constant, that is, motion takes place in a plane
exactly as in the Newtonian central force problem. By suitable orientation
of axes, we can set $\theta =\pi /2$. Next, since the metric tensor $g_{\nu
\lambda }$ do not depend on $t$ and $\varphi $, the corresponding components
of $U_{\mu }$ must be the constants of motion. These constants are $U_{0}$
and $U_{3}$ given by 
\begin{equation}
U_{0}=g_{00}U^{0}=B(r)\frac{dt}{dp},  \tag{A3}
\end{equation}%
\begin{equation}
U_{3}=g_{33}U^{3}=-r^{2}\frac{d\varphi }{dp}.  \tag{A4}
\end{equation}%
The remaining constant is just the rest mass $m_{0}$. Then from (A2), we
obtain%
\begin{equation}
\left( \frac{dr}{dp}\right) ^{2}=U_{0}^{2}-U_{3}^{2}\frac{B(r)}{r^{2}}%
-m_{0}^{2}B(r).  \tag{A5}
\end{equation}%
Using (A3), we can rewrite (A5) as%
\begin{equation}
\left( \frac{dr}{dt}\right) ^{2}=B^{2}(r)-\left( \frac{U_{3}}{U_{0}}\right)
^{2}\frac{B^{3}(r)}{r^{2}}-\frac{m_{0}^{2}}{U_{0}^{2}}B^{3}(r).  \tag{A6}
\end{equation}%
From (A3) and (A4), we get%
\begin{equation}
\frac{d\varphi }{dt}=-\frac{U_{3}}{U_{0}}\frac{B(r)}{r^{2}}.  \tag{A7}
\end{equation}%
Eliminating $t$ between (A6) and (A7), we obtain%
\begin{equation}
\left( \frac{dr}{d\varphi }\right) ^{2}=\left( \frac{U_{0}}{U_{3}}\right)
^{2}r^{4}-r^{2}B(r)-\frac{m_{0}^{2}}{U_{3}^{2}}r^{4}B(r).  \tag{A8}
\end{equation}%
Putting $B(r)=1-\frac{2M}{r}+\gamma r-kr^{2}$ and setting $u=1/r$, we can
rewrite (A8) as%
\begin{equation*}
\left( \frac{du}{d\varphi }\right) ^{2}=\left[ k-\frac{m_{0}^{2}}{U_{3}^{2}}%
+\left( \frac{U_{3}}{U_{0}}\right) ^{2}\right] -u^{2}+2Mu^{3}-u\gamma +\frac{%
2Mm_{0}^{2}u}{U_{3}^{2}}+\frac{km_{0}^{2}}{u^{2}U_{3}^{2}}-\frac{\gamma
m_{0}^{2}}{uU_{3}^{2}}.
\end{equation*}%
The term in square bracket on the right is constant. Differentiating both
sides of the above with respect to $\varphi $ and defining $h=\frac{U_{3}}{%
m_{0}}$, we get%
\begin{equation}
\frac{d^{2}u}{d\varphi ^{2}}=-u+3Mu^{2}-\frac{\gamma }{2}+\frac{M}{h^{2}}+%
\frac{1}{2h^{2}u^{2}}\left( \gamma -\frac{2k}{u}\right) ,  \tag{A9}
\end{equation}%
which is Eq.(2) in the text.

\begin{center}
\textbf{Appendix B}
\end{center}

It is useful to state the type of (in)stabilities of an equilibrium state we
had been discussing in the text. An equilibrium state corresponds to a
constant solution of a differential equation describing a mechanical system
and conversely. Constant solution means that velocity $\overset{.}{x}(=y)$
and acceleration $\overset{..}{x}$ be \textit{simultaneously} zero. The
concept of (in)stability of an equilibrium state is best understood from the
familiar example of motion of a pendulum about the equilibrium state $x=0$
and $\overset{.}{x}=0$, where $x$ is angle with the vertical. The motion is
stable because a small displacement from the equilibrium state will lead to
only small oscillations of the bob about the position ($0,0$) in a vertical
plane.

An equilibrium state in actual space is represented by a point on the phase
space ($x,\overset{.}{x}$). Different closed phase paths $y=f(x)$ around an
equilibrium point on the phase plane ($x,y$) correspond to oscillations of
the bob in the actual ($x,t$) space with different periods. In the dynamical
terminology, the point ($0,0$) on the phase plane is then called a stable 
\textit{center}. If a small displacement from any equilibrium state takes
the system far away from it, the state is represented by an unstable
equilibrium point on the phase plane. For instance, the equilibrium state $%
x=\pi $ and $\overset{.}{x}=0$, that is, the point ($\pi ,0$) on the phase
plane is an unstable \textit{saddle }point in the pendulum motion because
due to a small perturbation, the bob moves away from the point ($\pi ,0$)
never to return. This is reflected on the phase plane by open paths
originating from the equilibrium point. Finally, there could be whirling
motions of the bob off the vertical plane. This is picturized on the phase
plane by open paths never passing through an equilibrium point and there is
no question of stability here.

With the above understanding, the dynamical framework proceeds with
linearizing the general system [17]%
\begin{align}
\overset{.}{x}& =X(x,y)  \tag{B1} \\
\overset{.}{y}& =Y(x,y)  \tag{B2}
\end{align}%
around an equilibrium point ($x_{0},y_{0}$) such that 
\begin{eqnarray}
X(x_{0},y_{0}) &=&0,  \TCItag{B3} \\
Y(x_{0},y_{0}) &=&0.  \TCItag{B4}
\end{eqnarray}%
(Note that the family of phase paths $y=f(x)$ is obtained by integrating $%
\frac{dy}{dx}=\frac{Y}{X}$). Taylor expanding around ($x_{0},y_{0}$)
retaining only first order derivatives, we obtain from (B1), (B2) the linear
system%
\begin{equation}
\overset{.}{x}=a(x-x_{0})+b(y-y_{0})\text{, \ }\overset{.}{y}%
=c(x-x_{0})+d(y-y_{0})\text{,}  \tag{B5}
\end{equation}%
where%
\begin{equation}
\left[ 
\begin{array}{cc}
a & b \\ 
c & d%
\end{array}%
\right] =\left[ 
\begin{array}{cc}
\frac{\partial X}{\partial x} & \frac{\partial X}{\partial y} \\ 
\frac{\partial Y}{\partial x} & \frac{\partial Y}{\partial y}%
\end{array}%
\right] ,  \tag{B6}
\end{equation}%
the derivatives being calculated at ($x_{0},y_{0}$). Depending on the values
of $p,q$ and $\Delta $ defined by%
\begin{equation}
p=a+d\text{, \ \ }q=ad-bc\text{, \ }\Delta =p^{2}-4q,  \tag{B7}
\end{equation}%
one defines the stable center ($p=0,q>0,\Delta <0$), unstable saddle point ($%
q<0,\Delta >0$), stable node ($p<0,q>0,\Delta >0$), stable spiral ($%
p<0,q>0,\Delta <0$), unstable node ($p>0,q>0,\Delta >0$), unstable spiral ($%
p<0,q>0,\Delta <0$), degenerate stable node ($p<0,q>0,\Delta =0$),
degenerate unstable node ($p>0,q>0,\Delta =0$) etc. These classifications,
based on the eigen values of the matrix (B6), characterize on the phase
plane the paths in the neighborhood of ($x_{0},y_{0}$).

It is known that the characterization of a stable center via linear
approximation alone is `fragile' [17], hence unreliable, due to the fact
that it demands a rather stringent exact relation ($a+d=0$) to be fulfilled
by the first order derivatives. There are indeed specific problems where a
stable center defined in this way reflects only a \textit{transition}
between stable spiral and unstable node whereas the true behavior of the
full system (B1), (B2) is entirely different. \ A more reliable approach is
to consider the extrema of the Hamiltonian $H$ defining a center ($q_{0}>0$%
), as we did in the text. If we put, from Eqs.(12) and (13), the relations $%
\frac{\partial H}{\partial y}=X\left( x,y\right) $ and $\frac{\partial H}{%
\partial x}=-Y\left( x,y\right) $ both evaluated at ($x_{0},y_{0}$), into
(B7), we immediately see that $p=a+d=0$ identically. Also, the value of $q$
at ($x_{0},y_{0}$) becomes exactly what we designated in the text as $q_{0}$%
. Clearly, this is a second derivative approach, hence conclusive\footnote{%
One might imagine a loose analogy with notions in Einstein's theory of
general relativity. In this theory, locally flat inertial frames are defined
by the constraint that first order derivatives (analogous to $\overset{.}{x}$%
) of the metric tensor be zero at a small neighborhood of a point in that
frame. It does not however mean that the curvature at that point is zero
because the second order derivatives (analogous to $\overset{..}{x}$) are 
\textit{not} constrained to be zero. If they are also zero, then the
curvature vanishes and the space-time is truly flat there. So one could say
that curvature parallels instability while flatness parallels stability. }.
When $q_{0}=0$, the extrema of $H$ marginally corresponds to a center but
complicated saddles are not excluded. When $q_{0}<0$, the equilibrium point (%
$x_{0},y_{0}$) is definitely unstable, likely to be a saddle.

Observationally, we know that there are material orbits in the halo
reflecting the light coming to us and those orbits must be stable to
perturbations. Otherwise, such orbits would have collapsed to the galactic
center. Thus they should correspond to stable center representing periodic
motion in the real ($x,t$) plane, which is the physical criterion we used.
However, stable nodes or spirals on the phase plane are not ruled out 
\textit{in principle}, but none of them represents any periodic motion in
the real ($x,t$) plane as opposed to what is observed in the redshift
measurements. The speculation of huge amount of non-luminous dark matter
being hidden in the halo originates from the observed Doppler emissions from
stable circular orbits of neutral hydrogen clouds [2,5,10,11,19]. Therefore,
stable modes on the phase plane other than that of the stable center should
be ruled out on observational grounds.

\bigskip

\textbf{Figure captions:}

Fig.1. The upper line corresponds to $\gamma =7\times 10^{-28}$\ cm$^{-1}$\
and the lower line to $\gamma =-7\times 10^{-28}$\ cm$^{-1}$\ in the case of
Abell 2744. The plot remains essentially the same for other lenses tabulated
in Ref.[14].

Fig. 2. Curves I correspond to $\gamma =7\times 10^{-28}$\ cm$^{-1}$\ giving
a singularity of $q_{0}$\ at $R_{\text{sing}}=8.14\times 10^{28}$cm $>R_{%
\text{dS}}$.

\bigskip Fig. 3. Curves II correspond to $\gamma =-7\times 10^{-28}$\ cm$%
^{-1}$\ giving a singularity of $q_{0}$\ at $R_{\text{sing}}=9.11\times
10^{22}$cm $<R_{\text{E}}$.

\bigskip

\textbf{References}

[1] V. Sahni, \textquotedblleft Dark matter and dark
energy\textquotedblright\ [arXiv: astro-ph/0403324 v3]

[2] K. Lake, \textquotedblleft Galactic halos are Einstein clusters of
WIMPs\textquotedblright\ [arXiv: gr-qc/0607057 v3]

[3] J. D. Bekenstein and R. H. Sanders, Astrophys. J. \textbf{429}, 480
(1994), J. D. Bekenstein, Phys. Rev. D \textbf{70}, 083509 (2004); Erratum: 
\textit{ibid}. D \textbf{71}, 069901 (2005)

[4] M. Milgrom, Astrophys. J. \textbf{270}, 365 (1983); \textit{ibid.} 
\textbf{270}, 371 (1983); \textit{ibid.} \textbf{270}, 384 (1983); Phys.
Rev.D \textbf{80}, 123536 (2009)

[5] F. Rahaman, M. Kalam, A. DeBenedictis, A.A. Usmani and Saibal Ray, Mon.
Not. R. Astron. Soc. \textbf{389}, 27 (2008)

[6] T. Matos, F.S. Guzm\'{a}n and D. Nu\~{n}ez, Phys. Rev. D \textbf{62},
061301 (2000); K.K. Nandi, I. Valitov, and N.G. Migranov, Phys. Rev. D 
\textbf{80}, 047301 (2009). For the analysis of stability of scalar field
models, see, e.g., K.A. Bronnikov and S.V. Grinyok, Grav. \& Cosmol.\textbf{%
10}, 237 (2004); \textit{ibid.} \textbf{11}, 75 (2005); K.A. Bronnikov, M.V.
Skvortsova, and A.A. Starobinsky, Grav. Cosmol.\textbf{16}, 216 (2010); J.A.
Gonzalez, F.S. Guzman and O. Sarbach, Class. Quant. Grav.\textbf{\ 26},
015010 (2009); \textit{ibid}. \textbf{26}, 015011 (2009).

[7] P. D. Mannheim and D. Kazanas, Astrophys. J. \textbf{342}, 635 (1989);
P. D. Mannheim, Astrophys. J. \textbf{479}, 659 (1997); P. D. Mannheim,
Phys. Rev. D \textbf{75}, 124006 (2007). See also the review:\ P.D.
Mannheim, Prog. Part. and Nucl. Phys. \textbf{56}, 340 (2006)

[8] S. Pireaux, Class. Quant. Grav. \textbf{21}, 4317 (2004)

[9] A. Edery and M. B. Paranjape, Phys. Rev. D \textbf{58}, 024011 (1998)

[10] S. Bharadwaj and S. Kar, Phys. Rev. D \textbf{68}, 023516 (2003)

[11] U. Nucamendi, M. Salgado and D. Sudarsky, Phys. Rev. D \textbf{63},
125016 (2001)

[12] J.N. Islam, Phys. Lett. A \textbf{97}, 239 (1983)

[13] W. Rindler and M. Ishak, Phys. Rev. D \textbf{76}, 043006 (2007); See
also the recent invited review: M. Ishak and W. Rindler, "The Relevance of
the Cosmological Constant for Lensing" Gen. Rel. Grav. \textbf{42}, 2247
(2010) [arXiv:1006.0014 astro-ph CO].

[14] M. Ishak, W. Rindler, J. Dossett, J. Moldenhauer and C. Allen, Mon.
Not. Roy. Astron. Soc. \textbf{388}, 1279 (2008)

[15] Amrita Bhattacharya, Ruslan Isaev, Massimo Scalia, Carlo Cattani and
Kamal K. Nandi, JCAP \textbf{09} (2010) 004

[16] Amrita Bhattacharya, Guzel M. Garipova, Ettore Laserra, Arunava Bhadra
and Kamal K. Nandi, \textquotedblleft The vacuole model: new terms in the
second order deflection of Light\textquotedblright , JCAP \textbf{02} (2011)
028

[17] D.W. Jordan and P. Smith, \textit{Nonlinear ordinary differential
equations}, 3rd edition (Oxford University Press, Oxford, 1999)

[18] A. Palit, A. Panchenko, N.G. Migranov, A. Bhadra and K.K. Nandi, Int.
J. Theor. Phys. \textbf{48}, 1271 (2009)

[19] K. K. Nandi, A.I. Filippov, F. Rahaman, Saibal Ray, A. A. Usmani, M.
Kalam and A. DeBenedictis, Mon. Not. R. Astron. Soc. \textbf{399}, 2079
(2009)

[20] S.K. Bose, \textit{Introduction to general relativity} (Wiley Eastern,
New Delhi, 1980)

[21] S. Carloni, P. K. S. Dunsby, S. Capozziello and A. Troisi, Class.
Quant. Grav. \textbf{22}, 4839 (2005)

\bigskip

\end{document}